\documentclass{jetpl}
\usepackage[koi8-r]{inputenc}
\usepackage[russian]{babel}
\usepackage{color,amsmath,amsfonts,graphics,epsfig,amssymb}
\twocolumn
\lat

\title{%
\textit{Carpet--2} search for PeV gamma rays associated with IceCube
high-energy neutrino events }

\rtitle{Search for PeV gamma rays associated with IceCube events}

\sodtitle{
Carpet--2 search for PeV gamma rays associated with IceCube high-energy
neutrino events
}

\author{~\\[-7mm]
D.\,D.\,Dzhappuev$^{a}$,
I.\,M.\,Dzaparova$^{a,b}$,
E.\,A.\,Gorbacheva$^{a}$,
I.\,S.\,Karpikov$^{a}$,
M.\,M.\,Khadzhiev$^{a}$,
N.\,F.\,Klimenko$^{a}$,
A.\,U.\,Kudzhaev$^{a}$,
A.\,N.\,Kurenya$^{a}$,
A.\,S.\,Lidvansky$^{a}$,
O.\,I.\,Mikhailova$^{a}$,
V.\,B.\,Petkov$^{a,b}$,
K.\,V.\,Ptitsyna$^{a}$,
V.\,S.\,Romanenko$^{a}$,
G.\,I.\,Rubtsov$^{a}$,
S.\,V.\,Troitsky$^{a}$\thanks{Corresponding author; e-mail:
st@ms2.inr.ac.ru},
A.\,F.\,Yanin$^{a}$,
Ya.\,V.\,Zhezher$^{a}$\\
~}
\rauthor{Carpet--2 Group}

\sodauthor{D.\,D.\,Dzhappuev et al.\ (Carpet--2 Group)}

\address{
$^{a}$Institute for Nuclear Research of the Russian Academy of
Sciences,\\
60th October Anniversary prospect 7A, 117312 Moscow, Russia\\
$^{b}$Institute of Astronomy, Russian Academy of Sciences, Moscow, 119017
Russia}

\dates{December 06, 2018}{*}

\abstract{{\it Carpet--2}
is an air-shower array at Baksan Valley, Russia, equipped with a
large-area (175~m$^{2}$) muon detector, which makes it possible to separate
primary photons from hadrons. We report the first results of the search
for primary photons with energies $E_{\gamma}>1$~PeV, directionally
associated with IceCube high-energy neutrino events, in the data obtained
in 3080 days of {\it Carpet--2} live time.
}

\begin{document}
\maketitle
\section{Motivation}
\label{sec:intro}
The origin of high-energy ($E \gtrsim 100$~TeV) astrophysical neutrinos
detected in recent years by the IceCube experiment \cite{IC, IC-13, IC-14,
IC-A15mu, IC-15a-15bmu, IC-A16mu, IC-17a, IC-17b} remains one of the most
intriguing problems in modern astroparticle physics (see e.g.\
Ref.~\cite{rev} for a review). The arrival directions of these neutrinos
do not demonstrate any significant excess towards the Galactic disk, see
e.g.\  \cite{IC-17b, ST-gal}, as it would happen for most models of
Galactic sources, while serious tensions with Fermi-LAT observations of
the diffuse gamma-ray flux are present for extragalactic source models,
see discussions in Ref.~\cite{rev} and below. A recent observation of a
coincidence of one neutrino event with a blazar flare \cite{IC-flare} does
not add much to the picture: even if the association is physical,
non-observations of candidate sources for $\sim 40$ similar events in the
IceCube data set, together with constraints from diffuse gamma rays and
the lack of clustering of neutrino events, constrain the contribution of
similar blazars to $\lesssim 10\%$ of the total astrophysical neutrino
flux \cite{Murase7percent, newHalzen}. We therefore need additional
diagnostic tools to shed light on the origin of high-energy neutrino
events.

In most astrophysical scenarios which do not involve particle physics
beyond the Standard Model, high-energy neutrinos are produced in decays of
charged $\pi$ mesons. These mesons are, in turn, born in hadronic and
photohadronic processes, where produced $\pi^{\pm}$'s are always
accompanied by $\pi^{0}$'s. The latters decay immediately to photon pairs,
therefore providing for accompanying gamma-ray fluxes similar to those of
neutrinos. This simple scheme justifies the \textit{multimessenger
approach}: the fluxes of cosmic rays which launch the process, gamma rays
and neutrinos appear to be related in a straightforward way, so that their
joint observations constrain source models efficiently.

Energetic gamma rays produce electron-positron pairs when propagating
through the background radiation \cite{Nikishov}. The mean free path is
minimal for photons  with energies $E_{\gamma }\sim (0.1-10)$~PeV, similar
to those of IceCube neutrinos, because these gamma rays produce pairs on
the most abundant Cosmic Microwave Background photons. The mean free path
of gamma rays with these energies is of order of the size of a galaxy, and
therefore they become an important diagnostic tool for the origin of
neutrinos \cite{Murase-Fermi, Murase-gamma, OK-ST-gamma}: accompanying
photons from Galactic sources arrive to the observer without a significant
attenuation, while those from beyond the Galaxy lose their energy in the
pair-production processes. In the latter case, resulting electrons and
positrons launch electromagnetic cascades of inverse Compton scattering
and further pair production, in which the entire energy of the initial
$\sim$PeV gamma ray is transformed into photons with $E_{\gamma }\sim
10$~GeV, for which the Universe is transparent. These photons contribute
to the extragalactic diffuse background measured by Fermi LAT
\cite{Fermi-diffuse}. Since their contribution cannot exceed the total
measured value of the diffuse flux, the energy emitted in $\sim$PeV
photons in the Universe is constrained. For the $\pi$-meson production
mechanism, this constrains also the energy in neutrinos. It appears that
the observed flux of IceCube neutrinos at energies below $E_{\nu }\sim
200$~TeV is in serious tensions with this constraints, if the standard
production mechanism in extragalactic sources is assumed (see e.g.\
Ref.~\cite{rev} for more details). This suggests that a significant part
of the astrophysical neutrinos observed by IceCube should come from our
Galaxy, and their accompanying (sub)PeV photons should arrive to the Earth
unattenuated. Indeed, some indications exist~\cite{Semikoz} for a turnover
in the Fermi-LAT diffuse gamma-ray spectrum at highest energies,
consistent with the corresponding unabsorbed component. However, the
highest-energy points in the spectrum correspond to a few TeV, that is
orders of magnitude lower than the energies studied at IceCube.

It is an experimentally complicated task to perform a wide-scale search
for higher-energy cosmic photons. Indeed, in the TeV domain, atmospheric
Cerenkov telescopes have quite narrow fields of view, while at even higher
energies, extensive air shower (EAS) arrays experience difficulties in
separation between primary photons and hadrons. Presently, very few
experiments in the world are sensitive to $\sim$PeV photons. The aim
of the present work is to report the results of the first multimessenger
study of IceCube neutrinos with PeV gamma rays.

\section{The Carpet--2
experiment as a PeV gamma-ray telescope}
\label{sec:experiment}
The results we report here have been obtained with the \textit{Carpet--2}
experiment at the Baksan Neutrino Observatory of INR RAS. A more detailed
description of the experiment may be found in Refs.~\cite{modernization,
0902.0252, Carpet-2, 1511.09397}. \textit{Carpet--2} is an EAS array
constructed and developed for simultaneous measurements of the
electromagnetic, muonic and hadronic components of air showers induced by
cosmic particles with primary energies above 50~TeV. It consists of the
central unit (400 individual liquid-scintillator detector stations forming
a continuous ``carpet'' of the total area of 200~m$^{2}$), 6 remote
stations (scintillator detectors of 9~m$^{2}$ area in each station,
separated by 30 to 40~m from the center of the main unit), neutron
detectors and a large-area underground muon detector (MD), which is the
key element for the present study. In the configuration used for the
present analysis, the area of MD was 175~m$^{2}$. It consists of a
continuous array of individual plastic-scintillator detectors located in
the underground tunnel covered by 2.5~m of soil. This corresponds to the
detection threshold of 1~GeV for vertical muons. The arrival directions of
air showers are determined with the help of remote detectors. For the
analysis, showers with reconstructed axes inside the central unit
(excluding border detectors) were selected. The shower size $N_{e}$ was
determined by fitting the particle densities measured by individual
detector stations in the central unit to the Nishimura--Kamata--Greisen
(NKG) \cite{NKG1, NKG2} profile with the fixed shower age $s=1$,
\[
f(r)=\frac{5}{4\pi}\frac{N_{e}}{R_{M}^{2}}
\left(\frac{r}{R_{M}} \right)^{-1}
\left(1+\frac{r}{R_{M}}   \right)^{-3.5},
\]
where $r$ is the distance between the measurement point and the shower
axis and $R_{M}\approx 95$~m is the Moliere radius. The observable of MD
is the total number of muons recorded in the 175~m$^{2}$ detector,
$n_{\mu}$. For the purposes of the present work, therefore, each event is
characterised by the arrival direction (the azimuthal, $\phi$, and zenith,
$\theta$, angles), the arrival time, $N_{e}$ and $n_{\mu}$. Events
satisfying the following criteria were included in the data set:

--- the number of central-unit detectors with nonzero readings is $\ge
300$;

--- the total signal in the central unit is $\ge 10^{4}$ relativistic
particles;

--- $1<n_{\mu}<250$.

The latter cut deserves some comments. The installation was initially
constructed for the study of the structure of hadronic EASs near the core
and not for gamma-ray astronomy. Therefore, for the main part of the data
set, the condition $n_{\mu}>1$ was included as a trigger requirement. As
we will see immediately, this is not the optimal cut for gamma-ray
searches as at low energies, many gamma-ray events do not satisfy it. It
is this cut which determines the lower energy threshold of
$E_{\gamma}=1$~PeV used in the present analysis. The cut $n_{\mu}<250$ is
determined by the detector saturation.

\begin{figure}
\centering
\includegraphics[width=0.96\columnwidth,clip]{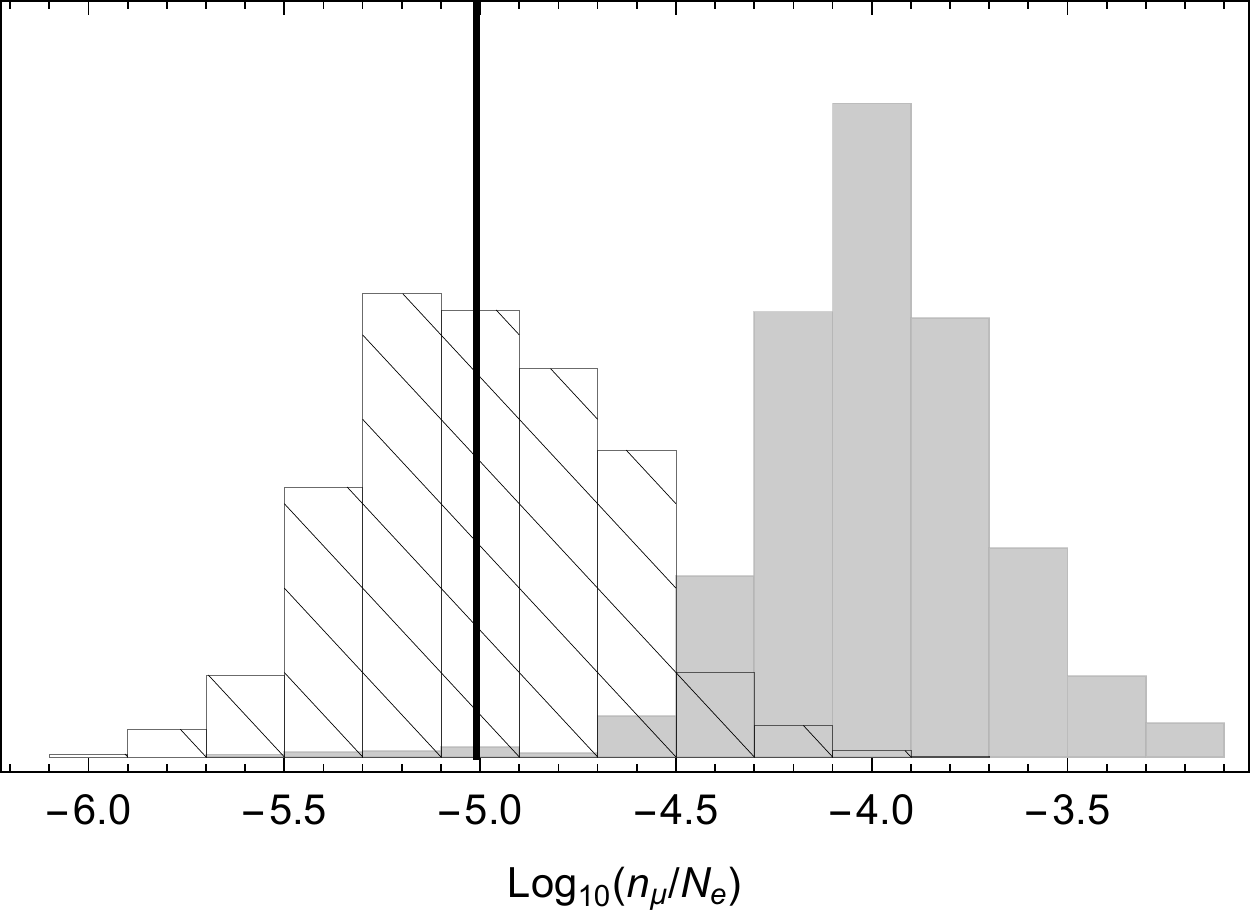}
\caption{\label{fig:distrib}
Figure~\ref{fig:distrib}:
Separation of primary photons from hadrons in terms of
$n_{\mu}/N_{e}$: distributions for MC events with primary photons
(hatched) and protons (shaded). The thick vertical line denotes the photon
median $C$. }
\end{figure}

Muons in EAS are produced dominantly in decays of charged mesons born in
hadronic interactions. Gamma-ray induced showers are mostly
electromagnetic and therefore are muon-poor compared to hadronic ones
\cite{MazeZaw}. We use this feature to search for candidate events with
gamma-ray primaries in the data. To determine the cuts separating primary
photons and hadrons, we perform Monte-Carlo (MC) simulations of air
showers based on
the CORSIKA~7.4003 \cite{CORSIKA} simulation
package with QGSJET-01c \cite{QGSJET-01}
and FLUKA2011.2c \cite{FLUKA1} as hadronic-interaction
models. We note in passing that for gamma rays of $\sim$PeV energies, the
description of the shower development is insensitive to the choice of the
hadronic model. The spectrum for primary photons $\propto
E_{\gamma}^{-2}$ was assumed while for hadrons, the spectrum was tuned
to reproduce the observed $N_{e}$ distribution. The arrival directions
were assumed to be isotropic; the simulations have been performed without
thinning. For each artificial shower, random positions of the axis in the
installation were assumed and the response of individual detectors was
simulated with a dedicated MC code. Air-shower parameters ($\phi$,
$\theta$, $N_{e}$, $n_{\mu}$) were subsequently reconstructed in a way
similar to the real data. Figure~\ref{fig:distrib}
presents the distributions of the ratios $n_{\mu}/N_{e}$
obtained in the MC simulations for primary photons and protons. We see
that they are indeed well separated, though some overlap is present.
Therefore, we determine the following criteria (inspired e.g.\ by
Ref.~\cite{0712.1147}) for the selection of photon candidates:

(1)~$N_{e} \ge N_{e}^{0}$, where $N_{e}^{0}=10^{5.03}$ is chosen from the
condition that 95\% of artificial photon showers with thrown primary
energy $E_{\gamma} \ge 1$~PeV satisfy this criterion;

(2)~$n_{\mu}/N_{e} \le C$, where $C=10^{-5.01}$ determines the ``photon
median'': 50\% of artificial photon showers satisfy this criterion.

In this way, a part of potential gamma-ray events is lost, and we account
for this in the efficiency calculation.

The reconstruction efficiency of the installation for primary photons is
not 100\%; it depends strongly on the primary energy and the zenith angle.
The total efficiency for $E_{\gamma} \ge 1$~PeV photons, as determined
from the MC simulations, is only 16\% but it grows rapidly with energy.
The $\theta$ dependence of the efficiency was derived from the real
distribution of zenith angles of photon candidate events. The overall
efficiency $\epsilon(\theta)$, obtained in this way, determines the
exposure for a particular source through the integration over the
observation time $t$,
\[
S\int\!dt\, \epsilon\left(\theta(t) \right)\cos\theta(t),
\]
where $S=162$~m$^{2}$ is the collecting area of the central unit.

\section{Data sets, analysis and results}
\label{sec:anal}
\paragraph{IceCube data.}
%\textbf{IceCube data.}
The main part of the study is the search for PeV
photons from a stacked sample of hypothetical sources of IceCube neutrino
events. We use all published IceCube arrival directions of high-energy
starting (HESE, Refs.~\cite{IC-14, IC-15a-15bmu, IC-17b}) and muon-track
\cite{IC-A15mu, IC-15a-15bmu, IC-A16mu, IC-17a} events, supplemented by
later online alerts \cite{alerts}. A useful
compilation of events is given in Ref.~\cite{Resconi}. We impose a
constraint that the arrival direction is
determined with a precision of better than 3$^{\circ}$ and is in the part
of the sky which {\it Carpet--2} can see (limited by $\theta \le
40^{\circ}$). The catalog of resulting events is given in
Table~\ref{tab:IC}.
\begin{table}
\begin{center}
\begin{tabular}{llll}
\hline
\hline
ID    &  R.A.   & DEC    & Error \\
\hline
HES13 &  67.9   & +40.3  & 1.2   \\
HES38 &  93.34  & +13.98 & 1.2   \\
HES47 &  209.36 & +67.38 & 1.2   \\
HES62 &  187.9  & +13.3  & 1.3   \\
HES63 &  160.0  & +6.5   & 1.2   \\
HES82 &  240.9  & +9.4   & 1.2   \\
DIF2  &  298.21 & +11.74 & 0.45  \\
DIF4  &  141.25 & +47.80 & 0.43  \\
DIF5  &  306.96 & +21.00 & 2.13  \\
DIF7  &  266.29 & +13.40 & 0.54  \\
DIF8  &  331.08 & +11.09 & 0.55  \\
DIF10 &  285.95 & +3.15  & 1.09  \\
DIF12 &  235.13 & +20.30 & 1.71  \\
DIF13 &  272.22 & +35.55 & 0.85  \\
DIF16 &  36.65  & +19.10 & 1.96  \\
DIF17 &  198.74 & +31.96 & 0.96  \\
DIF20 &  169.61 & +28.04 & 0.85  \\
DIF23 &  32.94  & +10.22 & 0.52  \\
DIF24 &  293.29 & +32.82 & 0.56  \\
DIF25 &  349.39 & +18.05 & 2.70  \\
DIF27 &  110.63 & +11.42 & 0.37  \\
DIF28 &  100.48 & +4.56  & 1.08  \\
DIF29 &  91.60  & +12.18 & 0.40  \\
DIF30 &  325.5  & +26.1  & 1.62  \\
DIF31 &  328.4  & +06.00 & 0.55  \\
DIF32 &  134.0  & +28.00 & 0.45  \\
DIF33 &  197.6  & +19.9  & 2.33  \\
DIF34 &  76.3   & +12.6  & 0.66  \\
DIF35 &  15.6   & +15.6  & 0.53  \\
EHE3  &  46.58  & +14.98 & 0.78  \\
EHE5  &  77.43  & +5.72  & 0.83  \\
EHE6  &  340.0  & +7.40  & 0.47  \\
AHES1 &  240.57 & +9.34  & 0.60  \\
AHES4 &  40.83  & +12.56 & 0.88  \\
\hline
\hline
\end{tabular}
\end{center}
\caption{\label{tab:IC}
Table~\ref{tab:IC}:
The catalog of 34 IceCube events used for the analysis. ID corresponds to
Ref.~\cite{Resconi}, R.A.\ and DEC are equatorial coordinates in degrees.
The last column gives the uncertainty in the arrival direction (in
degrees).}
\end{table}

\begin{figure}
\centering
\includegraphics[width=0.96\columnwidth,clip]{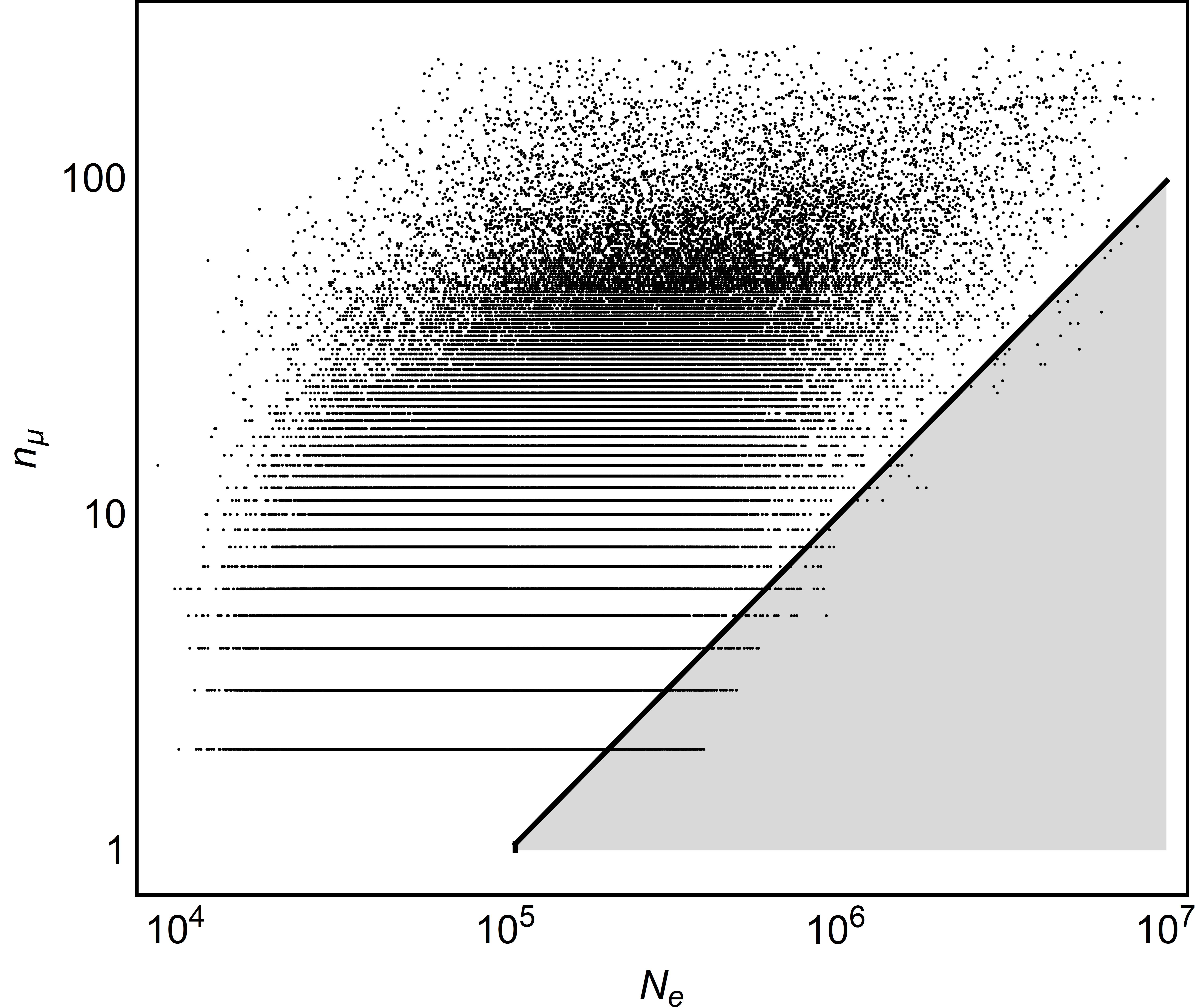}
\caption{\label{fig:cand-cuts}
Figure~\ref{fig:cand-cuts}:
Photon candidate selection on the $N_{e} - n_{\mu}$
plane. Dots: all events in the sample. Shaded region: photon candidates
(see text for details).
}
\end{figure}

\paragraph{\textit{Carpet--2} data.}
%\textbf{\textit{Carpet--2} data.}
We use 3080 live days of \textit{Carpet--2} data recorded between 1999 and
2011, in total 115821 events passing the cuts described above. The photon
candidate criteria select 523 events which we use for the stacked-source
analysis, see Fig.~\ref{fig:cand-cuts}.
In addition, in 2016--17, the
installation was operating for some periods of time in between works on
its upgrade (also limited by the damage from a natural disaster of
September 1, 2017); however, no useful MD data were recorded for this
time. Since April 7, 2018, \textit{Carpet--2} is working with the new
trigger optimized for gamma-ray searches (the $n_{\mu}>1$ condition is no
longer applied). All data of 2016--2018 were used to search for
coincidences with IceCube alerts, but not for the stacked search.

\paragraph{Stacked IceCube directions.}
%\textbf{Stacked IceCube directions.}
First, we determine the angular
resolution of the installation from MC simulations: 68\% of $E_{\gamma}
\ge 1$~PeV photons are reconstructed within the cone of $1.8^{\circ}$
around the thrown primary arrival direction. Taking into account the mean
uncertainty of the IceCube directions selected for the analysis and
following Ref.~\cite{BL-expectations}, we determine the optimal cut for
searching for directional correlations between \textit{Carpet--2} and
IceCube events as $3.0^{\circ}$ separation. We expect that 90\% of
potential gamma rays from sources would be reconstructed with this cut
and correct for 10\% of lost events in the exposure. The number of
Carpet--2 photon candidates within $3^{\circ}$ of the 34 IceCube events is
10. Next, we simulate isotropic arrival directions in the sky and filter
them with the exposure, $\epsilon(\theta)$, obtained in
Sec.~\ref{sec:experiment} , to find that the average number of random
coincidences between the two catalogs is 13.6. This corresponds to the
Poisson 95\% CL upper limit on the number of signal events $n_{95}<3.3$,
which transforms into \textbf{the 95\% CL limit on the steady flux of
$E_{\gamma} \ge 1$~PeV photons from the stacked arrival directions of 34
IceCube events of $<1.06 \times 10^{-14}$~cm$^{-2}$s$^{-1}$.}

\paragraph{Temporal correlations.}
%\textbf{Temporal correlations.}
The dates of the \textit{Carpet--2} main data set (1999--2011) do not
overlap with the IceCube working period. However, we use the events
recorded in the 2016--18 runs to search for coincidences with IceCube
events both in direction and time. The arrival direction of only one event
from the IceCube catalog was within the field of view of
\textit{Carpet--2} when the installation was taking data, the EHE3 muon
track recorded on December 10, 2016, at $20^{\rm h}07^{\rm m}16^{\rm
s}$~UT. The extimated neutrino energy was $\approx 100$~TeV. No events
with $N_{e} \ge N_{e}^{0}$ have been detected by \textit{Carpet--2} within
$3^{\circ}$ from the neutrino direction during December 9--11. From
simulations, 0.02 events were expected from a random coincidence. This
limits $n_{95}<3.0$ and allows us to constrain \textbf{the total fluence
of the potential flaring source of this neutrino event in $E_{\gamma} \ge
1$~PeV photons as $<5.4 \times 10^{-5}$~PeV/cm$^{2}$ (95\% CL).}

\section{Conclusions and outlook}
\label{sec:concl}
We have presented the first ever limits on PeV photons coming from the
arrival directions of IceCube high-energy neutrino events. These limits
may be used to constrain potential models of the neutrino origin in
Galactic point sources. Alternatively, if the extragalactic origin of
neutrinos is independently assumed, these results could constrain
new-physics models affecting gamma-ray propagation (see e.g.\
Ref.~\cite{ST-axion-rev} for a review), as it was suggested in
Ref.~\cite{Meyer-IC}.

The lack of Galactic-disk excess in IceCube data constrains the potential
contribution of known Galactic point sources, and purely diffuse models of
neutrino production in either the Galactic halo \cite{Aha-halo,
OK-ST-halo}
or in local sources~\cite{Semikoz, Semikoz-bubble} were put forward in
order to reconcile Fermi-LAT limits with the lack of disk anisotropy.
These models will be tested by the search of energetic diffuse gamma rays.
While \textit{Carpet-2} constraints on the diffuse gamma rays will be
reported elsewhere, we note that the best sensitivity will be achieved
with the upgraded \textit{Carpet--3} installation with the muon detector
of 410~m$^{2}$ and additional surface detector stations, both enlarging
the effective area of the installation and improving drastically the
gamma-hadron separation. \textit{Carpet--3} is assembled and will start
data taking in 2019. At the same time, the data recorded with the new,
``photon-friendly'', trigger are being taken now and will be used to lower
the threshold of the gamma-ray searches from 1~PeV to $\sim 100$~TeV
primary energy.

\section*{Acknowledgements}
Experimental work of the \textit{Carpet--2} installation is performed in
the laboratory of \textit{Unique Scientific Installation -- Baksan
Underground Scintillating Telescope} at the \textit{Collective Usage Center
``Baksan Neutrino Observatory of INR RAS''} under support of the Program of
fundamental scientific research of the RAS Presidium on ``Physics of
fundamental interactions and nuclear technologies''.
The work of a part of the group (DD, EG, MKh, NK, AUK, ANK, AL, OM, KP,
AY) on the upgrade of the installation was supported in part by the RFBR
grant 16-29-13049. The work of DD, ID, AUK, ANK, AL and VP was supported
in part by the RFBR grant 16-02-00687. ST thanks Olga Troitskaya for her
help in compiling the IceCube catalog and Michael Kachelriess, Oleg
Kalashev and Dmitri Semikoz for interesting discussions. Computer
calculations have been performed, in part, at the cluster of INR
Theoretical Physics Department.

\end{document}